# On The Double And Triple-Humped Fission Barriers And Half-Lives Of Actinide Elements

G. Royer, C. Bonilla

*Laboratoire Subatech, UMR: IN2P3/CNRS-Université-Ecole des Mines, 4 rue A. Kastler, 44307 Nantes Cedex 03, France*

**Abstract.** The potential energy of a deformed nucleus has been determined within a Generalized Liquid Drop Model taking into account the proximity energy, the microscopic corrections and compact and necked shapes. Multiple-humped potential barriers appear. A third minimum and third maximum exist in specific exit channels where one fragment is close to a magic spherical nucleus while the other one varies from oblate to prolate shapes. The heights of the fission barriers and half-lives of actinides are in agreement with the experimental results.

**Keywords:** Fission, Actinides, Liquid Drop Model, Half-lives.

**PACS:** 24.75.+i, 21.60.Ev, 27.90.+b.

## INTRODUCTION

The fission probability, the angular distribution of the fission fragments and the low energy $\alpha$ decay in some actinides support the hypothesis of hyperdeformed states lodging in a third well in several Th and U isotopes [1] confirming the pioneering work of Blons et al [2]. It is even also advocated that this third minimum could be the true ground state of the heaviest elements [3]. The potential barriers governing the actinide fission have been determined [4] within a Generalized Liquid Drop Model taking into account both the proximity energy when a neck exists, an accurate nuclear radius, the mass asymmetry and the microscopic corrections. The path leading rapidly to the formation of a deep neck in compact shapes has been selected and the ellipsoidal deformations of the separated fragments have been taken into account.

## FISSION BARRIERS

The proximity forces included in this GLDM allow to strongly lower the deformation energy in the quasi-molecular shape path leading rapidly to separated spherical fragments and allow to obtain the experimental fission barrier heights in the whole mass range, even for the Se, Br, Mo, In and Tb nuclei [5,6]. The $\alpha$ and cluster emission, the highly deformed rotating state [7] and fusion data can also be described within this unified approach.

In this work the coaxial ellipsoidal deformations have been taken into account since the limitation to spherical fragments leads to actinide fission barriers higher of some MeV than the experimental ones. The dependence of the potential barriers on the two-body shapes and microscopic corrections is displayed in Fig. 1. The shell effects generate the deformation of the ground state and increase the height of the first peak which appears already macroscopically. The proximity energy flattens the potential energy and will explain with the microscopic effects the formation of a second minimum lodging the superdeformed isomeric states for the heavier nuclei. In the two-sphere exit channel the rupture of the bridge of matter between the nascent fragments occurs before reaching the barrier top. The transition between one-body and two-body shapes is more sudden when the ellipsoidal deformations are allowed. It corresponds to the passage from a quasi-molecular one-body shape with spherical ends to two touching ellipsoidal fragments. The introduction of the microscopic energy still lowers the second peak ans shifts it to an inner position. It even leads to a third minimum and third peak. The heaviest fragment is a magic nucleus and remains almost spherical while the non magic fragment was born in an oblate shape. When the distance between mass centers increases the proximity energy tends to keep close the two tips of the fragments and the lighest one reaches a spherical shape which corresponds to a maximum of the shell energy, which is at the origin of the third peak. Later on, the proximity forces maintain in contact the fragments and the shape of the smallest one becomes prolate. Finally, a plateau exists at larger distances and much below the ground state when the proximity forces can no more compensate for the Coulomb repulsion and the fragments go away.

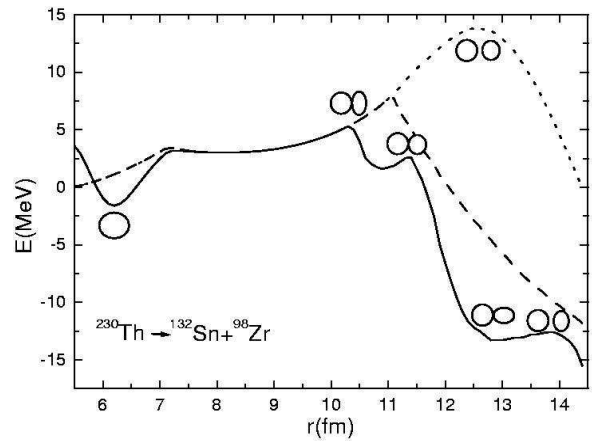

**FIGURE 1.** Fission barrier of a $^{230}$Th nucleus emitting a doubly magic $^{132}$Sn nucleus. The dotted and dashed lines correspond to the macroscopic energy within the two-sphere approximation and the ellipsoidal deformations for the two-body shapes. The solid line includes the microscopic corrections.

This third barrier appears only in the asymmetric decay channels and for some specific nuclei. In the symmetric mass exit path the proximity and Coulomb energies counterbalance the small shell effects and the two fragments remain in contact, one fragment being prolate while the other one is oblate before becoming both prolate at larger distances.

The whole reproduction of the heights of the inner and outer fission barriers which are almost constant from Th to Am isotopes is a very difficult task for all the theoretical approaches (liquid drop or droplet models, asymmetric two-center shell model, Hartree-Fock-Bogoliubov or Relativistic mean field theories,..). The theoretical and experimental energies of the maxima and minima of the potential barriers are compared in table 1. The choice of the most probable fission path is difficult for some elements since it exists a true degenerescence in energy between several mass asymmetry, particularly for the heaviest elements where the symmetric path seems more probable. The agreement with the experimental data is quite correct. For the heaviest nuclei the external barrier disappears.

**TABLE 1.** Experimental (e) and theoretical (t) first $E_a$, second $E_b$ and third $E_c$ peak heights and energies $E_2$ and $E_3$ of the second and third minima relatively to the ground state energy (in MeV).

| Reaction | $E_a(e)$ | $E_a(t)$ | $E_2(e)$ | $E_2(t)$ | $E_b(e)$ | $E_b(t)$ | $E_3(t)$ | $E_c(t)$ |
|---|---|---|---|---|---|---|---|---|
| $^{231}_{90}Th \rightarrow ^{132}_{50}Sn + ^{99}_{40}Zr$ | - | 5.5 | - | 5.2 | 6.5 | 7.1 | 3.9 / 5.6(e) | 6.9 / 6.3(e) |
| $^{233}_{90}Th \rightarrow ^{132}_{50}Sn + ^{101}_{40}Zr$ | - | 5.6 | - | 5.1 | 6.8 | 7.0 | 5.0 / 5.2(e) | 7.8 / 6.8(e) |
| $^{232}_{92}U \rightarrow ^{134}_{52}Te + ^{98}_{40}Zr$ | 4.9 | 4.5 | - | 3.2 | 5.4 | 5.0 | 4.2 | 5.1 |
| $^{234}_{92}U \rightarrow ^{131}_{50}Sn + ^{103}_{42}Mo$ | 5.6 | 5.0 | - | 4.4 | 5.5 | 5.9 | 3.7 / 3.1(e) | 5.6 |
| $^{235}_{92}U \rightarrow ^{131}_{50}Sn + ^{104}_{42}Mo$ | 5.7 | 5.7 | 2.5 | 4.9 | 5.8 | 6.6 | 5.4 | 6.9 |
| $^{236}_{92}U \rightarrow ^{132}_{50}Sn + ^{104}_{42}Mo$ | 5.6 | 5.5 | 2.3 | 4.8 | 5.5 | 6.2 | 3.1 / 3.1(e) | 4.4 |
| $^{237}_{92}U \rightarrow ^{132}_{50}Sn + ^{105}_{42}Mo$ | 6.1 | 6.1 | 2.5 | 5.3 | 5.9 | 6.5 | 3.6 | 6.2 |
| $^{238}_{92}U \rightarrow ^{132}_{50}Sn + ^{106}_{42}Mo$ | 5.7 | 5.5 | 2.6 | 4.5 | 5.7 | 5.6 | 4.1 | 5.6 |
| $^{238}_{94}Pu \rightarrow ^{130}_{50}Sn + ^{108}_{44}Ru$ | 5.6 | 5.2 | 2.7 | 3.6 | 5.0 | 4.5 | 3.2 | 3.6 |
| $^{239}_{94}Pu \rightarrow ^{130}_{50}Sn + ^{109}_{44}Ru$ | 6.2 | 5.8 | 2.6 | 4.1 | 5.5 | 5.0 | 4.1 | 5.6 |
| $^{240}_{94}Pu \rightarrow ^{130}_{50}Sn + ^{110}_{44}Ru$ | 5.7 | 5.3 | 2.4 | 3.3 | 5.1 | 4.6 | - | - |
| $^{241}_{94}Pu \rightarrow ^{131}_{50}Sn + ^{110}_{44}Ru$ | 6.0 | 6.1 | 1.9 | 4.4 | 5.5 | 5.6 | 5.1 | 5.5 |
| $^{243}_{94}Pu \rightarrow ^{132}_{50}Sn + ^{111}_{44}Ru$ | 5.9 | 6.3 | 1.7 | 4.6 | 5.4 | 5.2 | 3.2 | 4.6 |
| $^{242}_{95}Am \rightarrow ^{131}_{50}Sn + ^{111}_{45}Rh$ | 6.5 | 6.8 | 2.9 | 5.1 | 5.4 | 5.7 | 4.1 | 5.1 |
| $^{244}_{95}Am \rightarrow ^{132}_{50}Sn + ^{112}_{45}Rh$ | 6.3 | 7.0 | 2.8 | 5.3 | 5.4 | 5.7 | 2.4 | 4.2 |
| $^{243}_{96}Cm \rightarrow ^{130}_{50}Sn + ^{113}_{46}Pd$ | 6.4 | 6.0 | 1.9 | 3.6 | 4.2 | 4.2 | 2.4 | 2.7 |
| $^{245}_{96}Cm \rightarrow ^{130}_{50}Sn + ^{115}_{46}Pd$ | 6.2 | 6.0 | 2.1 | 3.1 | 4.8 | 3.7 | - | - |
| $^{248}_{96}Cm \rightarrow ^{130}_{50}Sn + ^{118}_{46}Pd$ | 5.7 | 5.3 | - | 2.0 | 4.6 | 3.0 | - | - |
| $^{250}_{97}Bk \rightarrow ^{130}_{50}Sn + ^{120}_{47}Ag$ | 6.1 | 6.4 | - | 2.6 | 4.1 | 3.7 | - | - |
| $^{250}_{98}Cf \rightarrow ^{125}_{49}In + ^{125}_{49}In$ | 5.6 | 4.9 | - | 0.1 | - | 1.7 | - | - |
| $^{256}_{99}Es \rightarrow ^{128}_{50}Sn + ^{128}_{49}In$ | 4.8 | 5.9 | - | 0.8 | - | 2.4 | - | - |
| $^{255}_{100}Fm \rightarrow ^{127}_{51}Sb + ^{128}_{49}In$ | 5.7 | 5.5 | - | 0.3 | - | 1.9 | - | - |

# HALF-LIVES

Within this asymmetric fission model the decay constant is simply the product of the assault frequency by the barrier penetrability. Our theoretical predictions are compared with the experimental data [8,9] in table 2. There is a correct agreement on 24 orders of magnitude, except for the lighest U isotopes.

**TABLE 2.** Experimental and theoretical spontaneous fission half-lives of actinide nuclei.

| Reaction | $T_{1/2,exp}(s)$ | $T_{1/2,th}(s)$ |
|---|---|---|
| $^{232}_{92}U \rightarrow ^{134}_{52}Te + ^{98}_{40}Zr$ | $2.5 \times 10^{21}$ | $3.6 \times 10^{16}$ |
| $^{234}_{92}U \rightarrow ^{131}_{50}Sn + ^{103}_{42}Mo$ | $4.7 \times 10^{23}$ | $8 \times 10^{19}$ |
| $^{235}_{92}U \rightarrow ^{131}_{50}Sn + ^{104}_{42}Mo$ | $3.1 \times 10^{26}$ | $7.7 \times 10^{23}$ |
| $^{236}_{92}U \rightarrow ^{132}_{50}Sn + ^{104}_{42}Mo$ | $7.8 \times 10^{23}$ | $1.0 \times 10^{22}$ |
| $^{238}_{92}U \rightarrow ^{132}_{50}Sn + ^{106}_{42}Mo$ | $2.6 \times 10^{23}$ | $5.3 \times 10^{22}$ |
| $^{238}_{94}Pu \rightarrow ^{130}_{50}Sn + ^{108}_{44}Ru$ | $1.5 \times 10^{18}$ | $2.6 \times 10^{19}$ |
| $^{239}_{94}Pu \rightarrow ^{130}_{50}Sn + ^{109}_{44}Ru$ | $2.5 \times 10^{23}$ | $4.8 \times 10^{22}$ |
| $^{240}_{94}Pu \rightarrow ^{130}_{50}Sn + ^{110}_{44}Ru$ | $3.7 \times 10^{18}$ | $4.8 \times 10^{19}$ |
| $^{243}_{95}Am \rightarrow ^{133}_{51}Sb + ^{110}_{44}Ru$ | $6.3 \times 10^{21}$ | $1.1 \times 10^{23}$ |
| $^{243}_{96}Cm \rightarrow ^{122}_{48}Cd + ^{121}_{48}Cd$ | $1.7 \times 10^{19}$ | $3 \times 10^{21}$ |
| $^{245}_{96}Cm \rightarrow ^{130}_{50}Sn + ^{115}_{46}Pd$ | $4.4 \times 10^{19}$ | $3 \times 10^{20}$ |
| $^{248}_{96}Cm \rightarrow ^{130}_{50}Sn + ^{118}_{46}Pd$ | $1.3 \times 10^{14}$ | $7.7 \times 10^{15}$ |
| $^{250}_{98}Cf \rightarrow ^{140}_{55}Cs + ^{110}_{43}Tc$ | $5.2 \times 10^{11}$ | $4.9 \times 10^{11}$ |
| $^{250}_{98}Cf \rightarrow ^{132}_{52}Te + ^{118}_{46}Pd$ | $5.2 \times 10^{11}$ | $1.2 \times 10^{10}$ |
| $^{255}_{99}Es \rightarrow ^{128}_{50}Sn + ^{127}_{49}In$ | $8.4 \times 10^{10}$ | $8 \times 10^{9}$ |
| $^{256}_{100}Fm \rightarrow ^{121}_{47}Ag + ^{135}_{53}I$ | $1.0 \times 10^{4}$ | $82$ |
| $^{256}_{102}No \rightarrow ^{128}_{51}Sb + ^{128}_{51}Sb$ | $110$ | $0.9 \times 10^{-2}$ |
| $^{256}_{102}No \rightarrow ^{116}_{46}Pd + ^{140}_{56}Ba$ | $110$ | $0.3 \times 10^{-1}$ |